\documentclass[aps,pra,twocolumn,superscriptaddress]{revtex4-2}

\usepackage{amsmath}
\usepackage{amssymb}
\usepackage{mathrsfs}
\usepackage{verbatim}
\usepackage{epsfig}
\usepackage{color}
\usepackage{soul}
\usepackage{algorithm}
\usepackage{algpseudocode}
\usepackage{setspace}
\usepackage[hidelinks,colorlinks=true,linkcolor=blue,citecolor=blue]{hyperref}
\usepackage{graphicx}

\graphicspath{{.}{./EPS/}}

\newcommand* {\bra}[1]{\ensuremath{\langle {#1} |}}
\newcommand* {\ket}[1]{\ensuremath{| {#1} \rangle}}

\newcommand{\ketbra}[2]{|{#1}\rangle\langle{#2}|}

\newcommand{\hbrho}{\boldsymbol{\rho}}
\newcommand{\hbsigma}{\boldsymbol{\sigma}}

\begin{document}
\title{Optimal two-qubit gates in recurrence protocols of entanglement purification}

\author{Francesco Preti}
\affiliation{Peter Gr\"unberg Institute (PGI-8), Forschungszentrum J\"ulich, D-52425 J\"ulich, Germany}
\affiliation{Institute for Theoretical Physics, University of Cologne, D-50937 Cologne, Germany}
\email{f.preti@fz-juelich.de}

\author{Tommaso Calarco}
\affiliation{Peter Gr\"unberg Institute (PGI-8), Forschungszentrum J\"ulich, D-52425 J\"ulich, Germany}
\affiliation{Institute for Theoretical Physics, University of Cologne, D-50937 Cologne, Germany}

\author{Juan Mauricio Torres}
\affiliation{Instituto de F\'{\i}sica, Benem\'{e}rita Universidad Aut\'{o}noma de Puebla, Apartado Postal J-48, Puebla 72570, Mexico}

\author{J\'ozsef Zsolt Bern\'ad}
\affiliation{Peter Gr\"unberg Institute (PGI-8), Forschungszentrum J\"ulich, D-52425 J\"ulich, Germany}
\email{j.bernad@fz-juelich.de}

\date{\today}

\begin{abstract}
We propose and investigate a method to optimize recurrence entanglement purification protocols. The approach is based on a numerical search in the whole set of $SU(4)$ matrices with the aid of a quasi-Newton algorithm. Our method evaluates average concurrences where the probabilistic occurrence of mixed entangled states is also taken into account. We show for certain families of states that optimal protocols are not necessarily achieved by bilaterally applied controlled-NOT gates. As we discover several optimal solutions, the proposed method offers some flexibility in experimental implementations of entanglement purification protocols and interesting perspectives in quantum information processing.   

\end{abstract}

\maketitle

\section{Introduction}

Entanglement is a key resource for
several tasks in
quantum information, quantum simulations \cite{Monroe}, computation \cite{Albash}, and communication \cite{Pirandola}. These tasks are based on the idea of creating networks of quantum systems, where the generation of maximally entangled sates
between qubits in spatially separated nodes is essential. These networks, which may consist of distant or nearby nodes, have been thoroughly investigated for efficient processing and transfer of quantum information \cite{Awschalom}. However, due to interactions with an uncontrollable environment, mixed or non-maximally entangled states are produced. To protect quantum information and to guarantee a high performance of its processing, one can use quantum error correction \cite{Peres, Devitt} or quantum teleportation in combination with entanglement purification \cite{Bennett1, Deutsch, Bennett2}. Quantum error correction is characterized by the quantum capacity of the transmission channel, which can be compared with the
  yield of entanglement purification for both one- 
  and two-way classical communication \cite{Bennett2}. The latter is the subject of this article. In particular, we consider a so-called recurrence protocol \cite{Dur}, an iterative approach, which operates in each purification step only on two identical copies of states.

In this paper, we discuss entanglement purification from the point of view of optimality. Recently, optimized entanglement purification has been investigated with the help of genetic algorithms \cite{Krastanov}, where 
the analytical and numerical studies are based on Werner states \cite{Werner}. In fact, also the first ever proposed protocols are based on either Werner \cite{Bennett1} or Bell diagonal states \cite{Deutsch}. It has been 
shown that $4$ or $12$ local random $SU(2)$ transformations can convert any state into a Bell diagonal or Werner state, respectively \cite{Bennett2}. We have already argued that these random local unitary transformations 
and states obtained in consequence are only useful
for grasping and understanding the complex task of entanglement purification because they may waste useful entanglement \cite{Torres}. Therefore, here, we develop a method for general states and demonstrate it for 
several simple examples including also the Werner state. Nonetheless, our motivation also lies in the fact that an experimental implementation may not have enough control over the generated mixed entangled states and thus more general, adaptive, and optimal entanglement purification strategies have to be made available. In this general context, we assume that an experiment can still guarantee identical copies of states before the protocol takes place. 

Our method is based on quasi-Monte Carlo numerical sampling of the states, which undergo the purification protocol, and then the concurrence \cite{Wooters} of the output states is integrated over an {\it a priori} probability distribution 
function. This results in an average two-qubit gate-dependent cost function for the whole sample of states. We employ a quasi-Newton algorithm \cite{Liu} to solve this non-linear optimization problem 
on the whole $SU(4)$ group. 
We focus on increasing entanglement in each step, therefore, the
obtained two-qubit gate is optimal in the sense that it achieves, on average, a higher increase of entanglement of an input family of states for a given purification step. This is beneficial in reducing the number of qubit pairs required to purify a single two-qubit state, as this number grows exponentially with the number of steps.
We discuss the performance of our method for several examples and compare with protocols based on one bilateral application of controlled-NOT (CNOT) gates, the paradigmatic two-qubit operation used in the seminal purification protocols \cite{Bennett1,Deutsch}.

The paper is organized as follows. In Sec.~\ref{Method} we introduce our
method and give some elementary examples to allow further acquaintance with the concept of the introduced cost function. 
In Sec.~\ref{Results} we demonstrate our approach for 
one two-parameter and four one-parameter family of states. Numerical and analytical results are presented for concurrences and success probabilities.  In Sec.~\ref{Conclusions}
we summarize and draw our conclusions. Some details supporting the main text are collected in the Appendix~\ref{AppendixA}.

\section{Method}
\label{Method}
In this section, we describe how the recurrence protocol is optimized with respect to an input family of quantum states.

\subsection{Entanglement purification protocol}\label{EntPurif}
Let us consider the product state of two-qubit pairs 
\begin{equation}
  \hbrho=\rho^{A_1,B_1} \otimes\rho^{A_2,B_2},
\label{eq:rhoAB4}
\end{equation}
where qubit components of each pair are assumed to be spatially separated at nearby or distant locations. These locations are labeled by $A$ and $B$. In an 
entanglement purification protocol, one performs local quantum operations, which may not involve just two-qubit gates \cite{Bernad1}. 
This is followed by measurements on one of the pairs
at both locations. A classical communication between $A$ and $B$ results in a qubit pair with a higher degree of entanglement. Both pairs are assumed to start in the same state $\rho$ 
and the degree of the entanglement is usually measured by the 
fidelity with respect to one of the Bell states
\begin{align}
  \ket{\Psi^\pm}=\tfrac{1}{\sqrt2}\left(\ket{01}\pm\ket{10}\right), \,
  \ket{\Phi^\pm}=\tfrac{1}{\sqrt2}\left(\ket{00}\pm \ket{11}\right).
  \label{eq:Bellstates}
\end{align}
However, these states can be subject to local unitary transformations, which can cause some technical difficulties, when one uses fidelity, i.e.~using fidelity as a cost function would force us to search also for additional local unitary operations in order to align the output state of the protocol with the Bell basis. 
Furthermore,
using a fidelity restricts the purification process to a particular basis. For instance, in Refs. \cite{Bennett1,Deutsch} a state can be purified only if it presents fidelity greater than $1/2$  with respect to any Bell state in Eq.~\eqref{eq:Bellstates}. In this setting a mixed state with fidelity close to one with respect to the maximally entangled state
$\ket{\Psi_{\rm M}}=\left(\ket{\Phi^-}+i\ket{\Phi^+}+i\ket{\Psi^-}+\ket{\Psi^+}\right)/2$ 
would not be purifiable, as $\ket{\Psi_{\rm M}}$  has a fidelity of $1/4$ with respect to any Bell state.
As we intend to 
analyze the entanglement purification in a very general setup, we require an entanglement measure that is invariant under local unitary transformations. Therefore, we turn to the concurrence as a measure 
of the attainability of a maximally entangled state \cite{Wooters}:
\begin{equation}\label{eq:concurrence}
 \mathcal{C}(\rho)=\max\{0,\lambda_1-\lambda_2-\lambda_3-\lambda_4\}.
\end{equation}
Here $\lambda_1, \lambda_2, \lambda_3, \lambda_4$ are the square roots of the non-negative eigenvalues of the non-Hermitian matrix
\begin{equation}
 \tilde \rho = \rho (\sigma_{y}\otimes\sigma_{y})\rho^{*}(\sigma_{y}\otimes\sigma_{y}), \nonumber
\end{equation}
where the asterisk is the complex conjugation in the standard basis and $\sigma_{y}$ is the Pauli matrix. It is worth noting that there are other possible entanglement measures, such as the entanglement formation or the relative entropy of entanglement \cite{Plenio}, but we do not consider them in this article, because the concurrence is, from a numerical point of view, a more tractable entanglement measure for two-qubit states.

\begin{figure}[t!]
    \includegraphics[width=.4\textwidth]{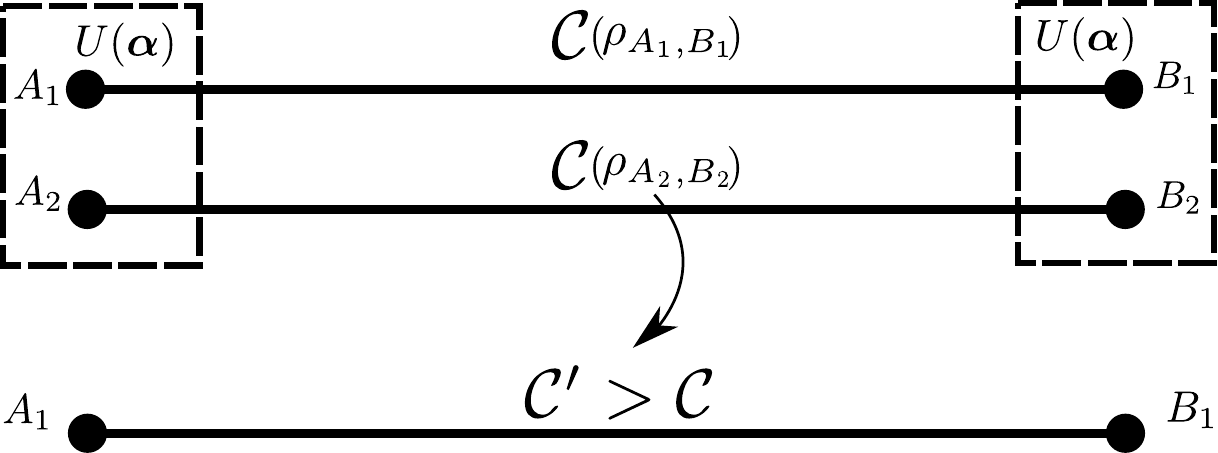}
    \caption{Schematic representation of bipartite entanglement purification, where the entanglement purification protocol trades two 
    entangled qubit pairs for a qubit pair with a higher degree of entanglement, which is quantified by the concurrence $\mathcal{C}$.}
    \label{fig:purifscheme}
\end{figure}

In this paper, we examine and optimize a purification protocol having the following steps.\\

(i) Two unitary transformations $\rho \rightarrow U(\boldsymbol{\alpha}) \rho U(\boldsymbol{\alpha})^\dagger$ are applied locally at $A$ and $B$ (see Fig.~\ref{fig:purifscheme}), where $U$ is a general two-qubit unitary described by a parameter vector $\boldsymbol{\alpha}$.
After the application of the quantum operation the four-qubit system 
attains the state  
\begin{equation}
\hbrho'=
 U_{A1,A2}(\boldsymbol{\alpha}) U_{B1,B2}(\boldsymbol{\alpha}) \hbrho U^\dagger_{B1,B2} (\boldsymbol{\alpha}) U^\dagger_{A1,A2} (\boldsymbol{\alpha}). 
\label{eq: maprho}
\end{equation}

(ii)  One of the pairs $(A_2,B_2)$ is then locally measured in the standard basis. There are four possible states, i.e., four-dimensional vectors, in which one can find the measured pair: 
\begin{eqnarray}
 \ket{1}&=&\ket{00}_{A_2,B_2}, \quad \ket{2}=\ket{01}_{A_2,B_2}, \nonumber \\
 \ket{3}&=&\ket{10}_{A_2,B_2}, \quad \ket{4}=\ket{11}_{A_2,B_2}. \nonumber
\end{eqnarray}
A successful measurement of one of the states $\ket{i}$ with $i\in\{1,2,3,4\}$ results in
a state for the other qubit
\begin{equation}
  \tilde\rho^{A_1,B_1}_i= \frac{\bra{i} \hbrho' \ket{i} }{\mathrm{Tr}\{\ketbra{i}{i} \hbrho'\}}
\label{}
\end{equation}
with probability
\begin{equation}
 p_i=\mathrm{Tr}\{\ketbra{i}{i} \hbrho'\}. \nonumber
\end{equation}

(iii) Depending on the value of the measurement results, which are communicated between the two parties, the state with the largest concurrence and the related success probability are kept, whereas the others are discarded. The output of the protocol is the pair 
$(\mathcal{C}',P)$, with the concurrence value $\mathcal{C}'$ of the state obtained with probability $P$. If there are multiple maxima, e.g.~, $\mathcal{C}(\tilde\rho^{A_1,B_1}_1)=\mathcal{C}(\tilde\rho^{A_1,B_1}_2)$, then 
$\mathcal{C}'=\mathcal{C}(\tilde\rho^{A_1,B_1}_1)$ and $P=p_1+p_2$.\\

Given two copies of a state $\rho$ with concurrence $\mathcal{C}$, it is straightforward to see that the pair $(\mathcal{C}',P)$ depends on the vector $\boldsymbol{\alpha}$, which we use as the optimization parameter.

Recurrence entanglement purification protocols might include symmetric and asymmetric single-qubit gates before and after the bilateral action of the two-qubit operations \cite{Bennett1,Deutsch}. These are important in an analytical approach to maintain the form of the state after each iteration, however, entanglement is not affected by this process. In this work we omit these specific single-qubit gates, as we are focused on increasing the value of the concurrence and not on the specific form of the state after each iteration step. We may instead consider general single-qubit unitaries as part of more general parametrizations. Furthermore, we stress again that our approach aims to increase the amount of entanglement between the qubits pairs regardless of the basis. This is an improvement concerning previous approaches based on seminal protocols \cite{Bennett1,Deutsch}
where a fidelity greater than $1/2$ with respect to a Bell state is needed for a working entanglement purification protocol.

\subsection{Protocol optimization}\label{ProtOpt}

In order to optimize the protocol described in Sec.~\ref{EntPurif}, we first define an appropriate parametrization for the two-qubit unitary transformations used in Eq.~\eqref{eq: maprho}. In principle, one should consider elements from $U(4)$, the group of $4 \times 4$ unitary matrices, which contains the subgroup $SU(4)$. However, $U(4)$ is the semi-direct product of $U(1)$ and $SU(4)$, where elements of $U(1)$ are rotations of the unit circle \cite{Baker}. Therefore,
choosing elements from $SU(4)$ in Eq.~\eqref{eq: maprho}
represents the most general unitary quantum operation involving two-qubit gates at locations $A$ and $B$.
Elements in $SU(4)$ can be parametrized as \cite{Tilma1}:
\begin{eqnarray} \label{eq: eulerU}
 U(\boldsymbol{\alpha})=e^{i \sigma_3 \alpha_1} e^{i \sigma_2 \alpha_2} e^{i \sigma_3 \alpha_3} e^{i \sigma_5 \alpha_4} e^{i \sigma_3 \alpha_5} e^{i \sigma_{10} \alpha_6} 
 e^{i \sigma_3 \alpha_7} e^{i \sigma_2 \alpha_8} \nonumber \\ 
 e^{i \sigma_{3} \alpha_9} e^{i \sigma_5 \alpha_{10}} e^{i \sigma_3 \alpha_{11}} e^{i \sigma_{2} \alpha_{12}} 
 e^{i \sigma_3 \alpha_{13}} e^{i \sigma_8 \alpha_{14}} e^{i \sigma_{15} \alpha_{15}}, \nonumber 
\end{eqnarray}
with $\boldsymbol{\alpha}=(\alpha_1, \alpha_2, \dots, \alpha_{15})^T \in \mathbb{R}^{15}$ ($T$ denotes the transposition), and
\begin{eqnarray}
 && 0 \leqslant \alpha_1, \alpha_3, \alpha_5, \alpha_7, \alpha_9, \alpha_{11}, \alpha_{13} \leqslant \pi, \nonumber \\
 &&  0 \leqslant \alpha_2, \alpha_4, \alpha_6, \alpha_8, \alpha_{10}, \alpha_{12} \leqslant \frac{\pi}{2} \nonumber \\
 && 0 \leqslant \alpha_{14} \leqslant \frac{\pi}{\sqrt{3}}, \quad 0 \leqslant \alpha_{15} \leqslant \frac{\pi}{\sqrt{6}}, \label{eq:anglecond}
\end{eqnarray}
where $\sigma_i,\ i=1, ..., 15$ form a Gell-Mann type basis of the Lie group $SU(4)$ (see the Appendix \ref{AppendixA}). This is called the Euler angle parametrization of $SU(4)$, which is sufficient
to represent every element of the Lie group. For example, the canonical parametrization $e^{i H}$, where $H$ is a $4 \times 4$ self-adjoint matrix with trace zero, is not minimal, because after exponentiation we may have multiple wrappings around the great circles of the $7$-sphere. 

Our aim is to increase the concurrence, which is a non-linear function of the state $\rho$ and the protocol's unitary matrix $U$. Furthermore, we have lower and upper bounds on $\boldsymbol{\alpha}$, as given in Eq.~\eqref{eq:anglecond}. We then consider a cost function $f:\mathbb{R}^{15} \rightarrow 
\mathbb{R}$ as
\begin{align}
    f(\boldsymbol{\alpha}) = 1 - \mathcal{C}'(\boldsymbol{\alpha}) \label{eq:f}
\end{align}
and our optimization problem is, to find local minimizers, e.g.~, a point $\boldsymbol{\alpha}^*$ such that
\begin{equation}
 f(\boldsymbol{\alpha}^*)\leqslant f(\boldsymbol{\alpha}) \nonumber
\end{equation}
for all $\boldsymbol{\alpha}$ in the Euclidean norm defined neighborhood of $\boldsymbol{\alpha}^*$. The gradient $\nabla f (\boldsymbol{\alpha})$ can be made available to us due to an automatic differentiation \cite{JAX}, so we 
implement for the optimization a quasi-Newton algorithm the so-called limited memory Broyden-Fletcher-Goldfarb-Shanno (L-BFGS) method \cite{Liu}. Here, we have a constrained optimization problem, because $\boldsymbol{\alpha}$ takes values in the hyperrectangle defined by Eq.~\eqref{eq:anglecond} and thus we use the L-BFGS approach of Ref. \cite{Ryrd}. 
As long as the state to be purified is given, the above approach yields an optimal $\boldsymbol{\alpha}^*$ and a concurrence $\mathcal{C}'$ as close as possible to one. 
The vector $\boldsymbol{\alpha}^*$ gives the best two-qubit gate associated with this given state for a given purification step. However, our main aim is to provide the best quantum gates, which are able to purify most effectively certain classes of states and not only a fixed one. This is relevant in a scenario where the generation of distant entanglement is affected by altering noise, leading to slightly different types of states entering the protocol.
In order to formulate this quantitatively we introduce  a probability density function (PDF) $p(\boldsymbol{x})$, where the vector $\boldsymbol{x}$ defines uniquely the state $\rho$ according to a parametrization. 
For example, in the case of a  Werner state \cite{Werner}
\begin{eqnarray}
\rho(x) &=& x \ket{\Psi^{-}}\bra{\Psi^{-}} + \frac{1-x}{3} \ket{\Psi^{+}}\bra{\Psi^{+}} \nonumber \\
&+& \frac{1-x}{3}  \ket{\Phi^{-}}\bra{\Phi^{-}}+ \frac{1-x}{3} \ket{\Phi^{+}}\bra{\Phi^{+}},  
\label{eq:wernerstate}
\end{eqnarray}  
we have $x\in[0,1]$ and the PDF satisfies
\begin{equation}
 \int^1_0 p(x)\, dx=1. \nonumber
\end{equation}
In general, a two-qubit state can be described by $15$ parameters, which have to fulfill some non-trivial conditions \cite{Kimura,Byrd,Gamel}. The choice of the PDF is not straightforward and the only guideline 
we have is that the support $\operatorname{supp}(p)$ consists of all $\boldsymbol{x}$, which define entangled states. This is motivated by 
the fact that separable states are not purifiable. In the case of the Werner states, the support of the PDF is the interval $(0.5,1]$. Thus, the PDF may put more weight on states with a given concurrence.
\begin{algorithm}[H]
		\setstretch{1.20}
		\caption{Optimization of recurrence protocol}
		\label{alg:ouralgo}
		\hspace*{\algorithmicindent} \textbf{Input} $\hbrho(\boldsymbol{x})$, optimizer $\text{OPT}$\\
		\hspace*{\algorithmicindent} \textbf{Output} $\hbrho(\boldsymbol{x})$
		\begin{algorithmic}[1]
		\For{$j=1$ to $M$}
		\State $\boldsymbol{x}_j \sim p(\boldsymbol{x}_j)$
		\State $\hbrho_j = \hbrho(\boldsymbol{x}_j) \otimes \hbrho(\boldsymbol{x}_j)$
		\EndFor
			\For{$N=1$ to $N_{\text{max}}$} \Comment{with random restart}
			\State $U_{AB}(\boldsymbol{\alpha}) = U_{A1,A2}(\boldsymbol{\alpha}) U_{B1,B2}(\boldsymbol{\alpha})$
			\For{$j=1$ to $M$}
			\State $\hbsigma^{j}(\boldsymbol{\alpha}) = U_{AB}(\boldsymbol{\alpha}) \hbrho_j U^{\dagger}_{AB}(\boldsymbol{\alpha})$
			\For{$l=1$ to $4$}
			\State $\hbsigma^{jl}_A(\boldsymbol{\alpha}) =  \frac{\bra{l} \hbsigma^{j}(\boldsymbol{\alpha})  \ket{l} }{\mathrm{Tr}\{\ketbra{l}{l} \hbsigma^{j}(\boldsymbol{\alpha})  \}}$
			\EndFor
			\EndFor
			\State $\mathcal{C}^l(\boldsymbol{\alpha}) = \frac{1}{M}\sum_{j=1}^M \mathcal{C}(\hbsigma^{jl}_A)$
			\State $l' = \underset{l=1,2,3,4}{\text{argmin}}\ \left[1 - \mathcal{C}^l(\boldsymbol{\alpha})\right]$
			\State $\bar{f}(\boldsymbol{\alpha}) =  1 - \mathcal{C}^{l'}(\boldsymbol{\alpha})$
			\State $\boldsymbol{\alpha}^* = \text{OPT}(\bar{f}(\boldsymbol{\alpha}), \nabla_{\boldsymbol{\alpha}} \bar{f}(\boldsymbol{\alpha}))$
			\For{$j=1$ to $M$}
			\State $\hbrho_j = \hbsigma^{jl'}_{A}(\boldsymbol{\alpha}^*) \otimes \hbsigma^{jl'}_{A}(\boldsymbol{\alpha}^*)$
			\EndFor
			\EndFor
		\end{algorithmic}
	\end{algorithm}
In this context, we have an output concurrence $\mathcal{C}'(\boldsymbol{\alpha}, \boldsymbol{x})$ depending on both the two-qubit gate and the input state. Therefore, we are going to use an average cost function 
\begin{equation}
 \bar{f}(\boldsymbol{\alpha}) = 1 - \int_{\operatorname{supp}(p)} \mathcal{C}'(\boldsymbol{\alpha}, \boldsymbol{x}) p(\boldsymbol{x})\,  d \boldsymbol{x} \label{eq:cost}
\end{equation}
in the L-BFGS algorithm. The integral can be either solved numerically or approximated via a quasi-Monte Carlo approach by sampling $\boldsymbol{x}$ over its corresponding parameter distribution $p(\boldsymbol{x})$:
\begin{align}
    \bar{f}(\boldsymbol{\alpha}) = 1 - \underset{\boldsymbol{x} \sim p(\boldsymbol{x})}{\mathbb{E}} \left[\mathcal{C}'(\boldsymbol{\alpha}, \boldsymbol{x}) \right].
\end{align}
We chose the second approach since it proved to be precise enough, i.e., 
\begin{equation}
\frac{1}{M} \sum^M_{j=1} p(x_j) \approx 1, \nonumber
\end{equation}
with $M$ being the sample size, and numerically faster. A summary of the optimization routine can be seen in Algorithm~\ref{alg:ouralgo}. We first sample a batch of random density matrices by sampling over the parameter space according to a probability distribution $p(\boldsymbol{x})$. Then we optimize the average output concurrence of the protocol as a function of the parameters $\boldsymbol{\alpha}$. The optimization routine provides us with output density matrices, which are re-inserted in the protocol for a successive purification round, controlled by a different unitary matrix $U(\boldsymbol{\alpha})$. This is also optimized, giving rise to a loop that breaks when the desired concurrence level or when the maximal number of iterations $N_{\text{max}}$ is reached. We would like to highlight that the unitary matrices which build the optimized purification protocol are different from each other. Each one of them is optimized for its own purification step, a procedure that can be conceived as a form of adaptive purification. Due to the nature of the optimization, and the intrinsic complexity of differentiating, e.g., the concurrence, it is likely that some of the concurrence values output by our optimization routine are not true optima, but rather local optima. However, in general, restarting the algorithm multiple times with different initial conditions can help reduce the probability of it being stuck in a local minimum.

Now, we shed light on the meaning of the average cost function through the following two examples. We employ the CNOT gate
\begin{equation}
 U_{\text{CNOT}}=\begin{pmatrix} 1 & 0 & 0 & 0 \\ 0 & 1 & 0 & 0 \\ 0 & 0 & 0 & 1\\ 0 & 0 & 1 & 0  \end{pmatrix}=
 e^{i\frac{3\pi}{4}} \times \underbrace{U'}_{\in SU(4)}, \nonumber
\end{equation}
where $U'$ in the Euler angle parametrization is given by setting
\begin{eqnarray}
 &&\alpha_3= \alpha_5=\alpha_7= \frac{\pi}{4}, \nonumber \\
 &&\alpha_4= \alpha_6=\alpha_{10}= \frac{\pi}{2}, \nonumber
\end{eqnarray}
and the remaining nine angles to zero in Eq.~\eqref{eq: eulerU}. By fixing the vector $\boldsymbol{\alpha}$, we are able to get a value for the average cost function of the CNOT gate and thus to evaluate its performance. 

We remark that 
for the following three examples, 
we merely calculate the cost function of the CNOT gate in order to explain this particular process,
and to show that this gate can be optimal in certain cases with a possible combination of other
single-qubit gates.
In Sec.~\ref{Results} we run an optimization process where the cost function is evaluated for many two-qubit gates in order to find the optimal one. 
\\

{\it Example II.1.}
Let us consider the state
\begin{equation}
 \frac{1}{6}\begin{pmatrix} 1+2x & 0 & 0 & 1-4x \\ 0 & 2-2x & 0 & 0 \\ 0 & 0 & 2-2x & 0\\ 1-4x & 0 & 0 & 1+2x  \end{pmatrix}, \quad \text{with} \quad x \in [0,1] \label{eq:example1}
\end{equation}
subject to the purification protocol with CNOT gates. This state is a rotated Werner state \cite{Werner} which was employed in the seminal protocol of Ref. \cite{Bennett1}
and its concurrence reads
\begin{equation}
 \mathcal{C}(x)=\begin{cases}2x-1, & x \in (0.5,1] \\ 0, & x \in [0, 0.5].\end{cases} \nonumber
\end{equation}
 The output reads
\begin{equation}
 \mathcal{C}'_{\text{CNOT}}(x)=\frac{3(4x^2-1)}{5-4x+8x^2} \quad \text{for} \quad x \in (0.5,1], \nonumber
\end{equation}
with success probability
\begin{equation}
 P_{\text{CNOT}}=\frac{5-4x+8x^2}{9}. \nonumber
\end{equation}
For the sake of simplicity, we consider a uniform PDF $p(x)$ with $\operatorname{supp}(p)=(0.5,1]$. Hence, the input average cost function reads
\begin{equation}
 \bar{f}_{\text{input}}=1-\int^1_{0.5} \mathcal{C}(x) p(x)\, dx =0.5 \nonumber
\end{equation}
and the application of the purification protocol with CNOT gates yields
\begin{equation}
 \bar{f}_{\text{CNOT}}=1-\int^1_{0.5} \mathcal{C}'(x) p(x)\, dx =0.450103. \nonumber
\end{equation}
The result shows that the protocol with two identical copies of states allows us to increase on average the entanglement of the output states.\\

{\it Example II.2.}
Now, we consider the state
\begin{equation}
 \begin{pmatrix} \frac{x}{2} & 0 & 0 & -\frac{x}{2} \\ 0 & 0 & 0 & 0 \\ 0 & 0 & 1-x & 0\\ -\frac{x}{2} & 0 & 0 & \frac{x}{2}  \end{pmatrix}, \quad \text{with} \quad x \in [0,1]. \label{eq:example2}
\end{equation}

Restricted to the interval $x\in [2/3,1]$ it corresponds to a maximally entangled mixed state
\cite{Ishizaka, Munro}. In general for every $x\in (0,1]$, this
 state with concurrence $\mathcal{C}(x)=x$ is perfectly purifiable in just one iteration of the protocol \cite{Torres}. We can find after one iteration that the concurrence becomes
\begin{equation}
 \mathcal{C}'_{\text{CNOT}}(x)=1, \nonumber
\end{equation}
with success probability
\begin{equation}
 P_{\text{CNOT}}=\frac{x^2}{2}. \nonumber
\end{equation}
It is immediate from Eq.~\eqref{eq:cost} that for {\it any} PDF with $\operatorname{supp}(p)=[0,1]$,
\begin{equation}
 \bar{f}_{\text{CNOT}}=1-\int^1_{0} \mathcal{C}'(x) p(x)\, dx =0. \nonumber
\end{equation}
This means that the CNOT gate is optimal for this family of states.

{\it Example II.3.}
As the last example let us consider the initial state
\begin{equation}
 x\ketbra{\Phi^+}{\Phi^+}+(1-x)\ketbra{\Phi^-}{\Phi^-}, \quad \text{with} \quad x\in[0,1],   
\end{equation}
and concurrence $\mathcal{C}(x)=|1-2x|$. After one iteration of the purification protocol with bilaterally applied CNOT gates, it is not hard to realize that the resulting state has the concurrence 
\begin{equation}
\mathcal{C}'_{\text{CNOT}}(x)=(1-2x)^2\nonumber 
\end{equation}
with success probability $P_{\text{CNOT}}=1$. The output
concurrence $\mathcal{C}'_{\text{CNOT}}(x)$ is less than or equal to $\mathcal{C}(x)$ for all $x \in [0,1]$. Thus, we conclude by using the properties of concurrence and integration that
\begin{equation}
 \int^1_{0} \mathcal{C}'_{\text{CNOT}}(x) p(x)\, dx \leqslant  \int^1_{0} \mathcal{C}(x) p(x)\, dx
\end{equation}
for any PDF with $\operatorname{supp}(p)=[0,1]$. Hence, the initial average cost function $\bar f_{\rm input}$ is always less than or equal to 
$\bar f_{\rm CNOT}$. This is an example where the purification fails with the sole implementation of the CNOT gate. It should be noted that
for  $x\neq 0.5$, the state in this example can be purified with previous protocols \cite{Bennett1,Deutsch} that work on the Bell basis, and where the implementation with the CNOT gate is now accompanied by local single-qubit gates.

These examples demonstrate the meaning of the average cost function. It is obvious that one or more two-qubit gates can be optimal for certain family of states and less optimal for others.
In the subsequent section, we will investigate numerically several cases. 

\begin{figure*}[t!]
 \begin{center}
 \includegraphics[width=.39\textwidth]{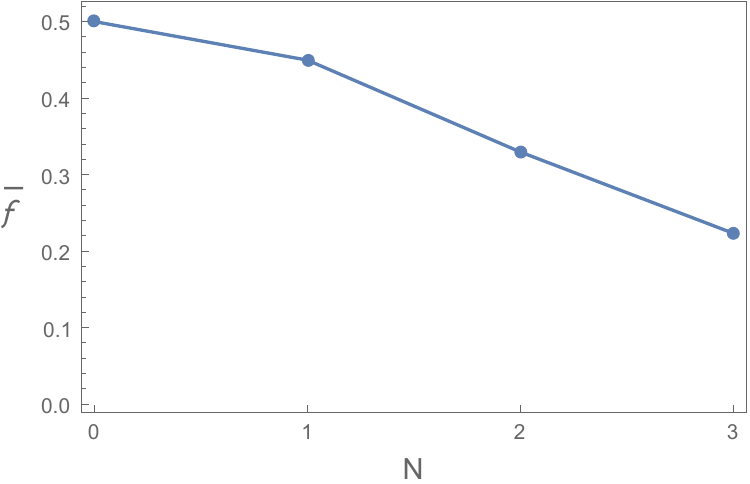}
 \includegraphics[width=.39\textwidth]{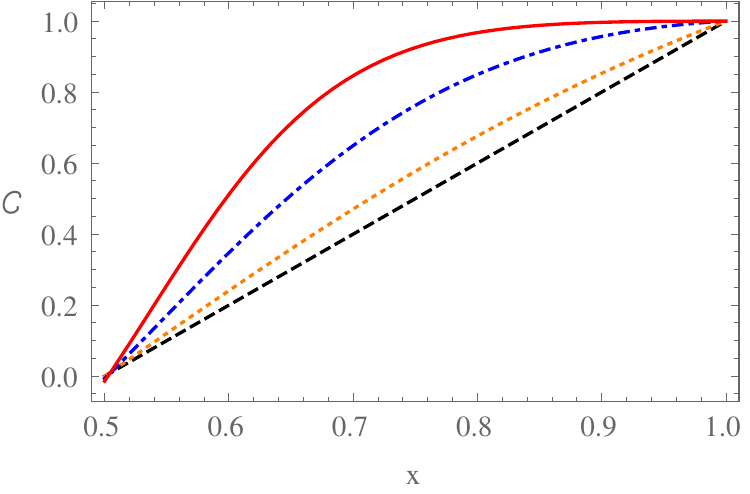}
 \includegraphics[width=.39\textwidth]{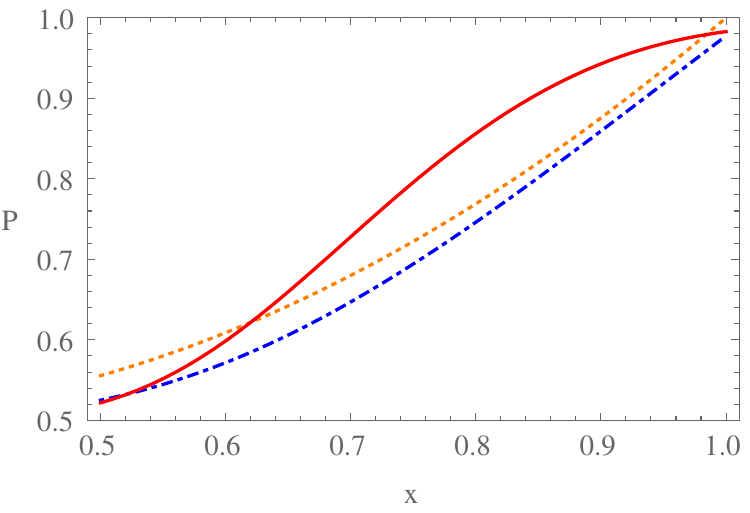}
 \includegraphics[width=.39\textwidth]{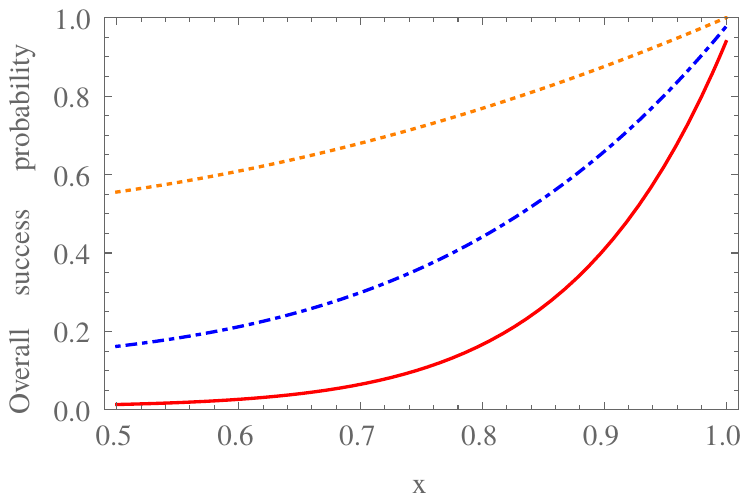}
\caption{Optimized purification protocol for the family of states in Eq.~\eqref{eq:example1}.
Top left panel: Average cost function $\bar{f}$ with a uniform PDF as a function of $N$, the number of iterations. Top right panel: Concurrence $\mathcal{C}$ as a function of $x$. 
Four curves are presented for different values of iterations $N$: $0$ or the input concurrence (black dashed line), $1$ (orange dotted line), $2$ (blue dash-dotted line), and $3$ (red solid line).
Bottom left panel: Success probabilities as a function of $x$ for different values of iterations $N$:  $1$ (orange dotted line), $2$ (blue dash-dotted line), and $3$ (red solid line). Bottom right panel: The overall success 
probability after $N=1$ (orange dotted line), $N=2$ (blue dash-dotted line), and $N=3$ (red solid line) iterations as a function of $x$. The sample size has been set to $M=1000$.}
   \label{fig:TWerner}
 \end{center}
\end{figure*}
\section{Results}
\label{Results}

In this section it is demonstrated how our proposed method can find optimal purification schemes. Our first case is the continuation of 
{\it Example II.1.} in Sec.~\ref{Method}. We have seen so far that 
the CNOT gate in $N=1$ purification round can reduce the average cost function $\bar{f}$ approximately by $0.05$. The simulation results with the input state in Eq.~\eqref{eq:example1} show that the optimal
$SU(4)$ gate for $N=1$ has a similar improvement on $\bar{f}$ as the CNOT gate (see Fig.~\ref{fig:TWerner}). The optimal gates found for $N=2$ and $3$ can further reduce $\bar{f}$; however, it is easy to check that 
the CNOT gate alone is not optimal anymore for these rounds of iterations. In Fig.~\ref{fig:TWerner} it is also shown that a higher number of iterations results in more concave shapes of the corresponding concurrences. It is 
worth noting that the success probability of the second iteration is lower than the success probability of the first iteration. The overall success probability of three iterations, 
displayed also in Fig.~\ref{fig:TWerner}, is defined as follows: In the first iteration four qubit pairs, in the second iteration two qubit pairs, and in the third iteration the final qubit pair are successfully purified. These 
results show that states close to maximally entangled states can be produced already in the third iteration, but the overall success probability of the procedure is getting smaller with the number of 
iterations. The only fixed point is  $\mathcal{C}=1$. Thus our method seems to provide the same success probabilities as the ones found in \cite{Bennett1}. As a result, we have investigated the circumstances where the CNOT gate is also optimal. It turns out that the local unitary transformation introduced by \cite{Deutsch} and given by
\begin{equation}
 b_{A_1}^{\dagger}\otimes b_{A_2}^{\dagger}\otimes b_{B_1}\otimes b_{B_2}, \nonumber
\end{equation}
where
\begin{equation}
b=\frac{\mathbb{I}_2+i\sigma_x}{\sqrt2}, \label{eq:localtr}
\end{equation}
with the Pauli matrix $\sigma_x$ and the identity map $\mathbb{I}_2$, is crucial for the CNOT gate. This is the transformation, which transforms a Werner state in Eq.~\eqref{eq:wernerstate} into 
the state in Eq.~\eqref{eq:example1}. Now if one applies Eq.~\eqref{eq:localtr} before all iterations of the protocol involving only the CNOT gate, then the same curves are obtained as in Fig.~\ref{fig:TWerner}. 
This means that there are more optimal protocols which yield the same results and our approach can find them.

\begin{figure*}[t!]
 \begin{center}
 \includegraphics[width=.39\textwidth]{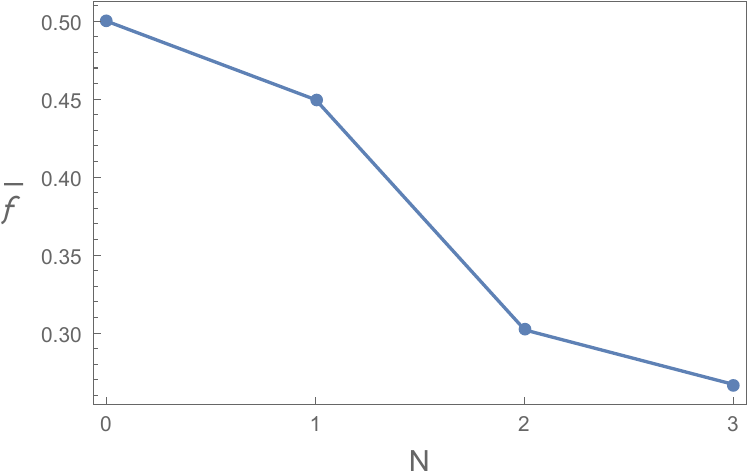}
 \includegraphics[width=.39\textwidth]{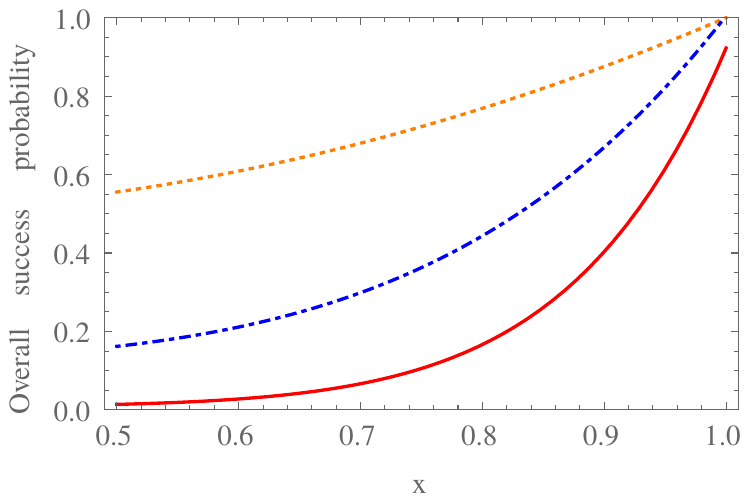}
\caption{Optimized purification protocol for Werner states, see Eq.~\eqref{eq:wernerstate}.
Left panel: Average cost function $\bar{f}$ with a uniform PDF as a function of $N$, the number of iterations.  Right panel: The overall success 
probability after $N=1$ (orange dotted line), $N=2$ (blue dash-dotted line), and $N=3$ (red solid line) iterations as a function of $x$. The sample size has been set to $M=1000$.}
   \label{fig:Werner}
 \end{center}
\end{figure*}

The one-parameter family of states in Eq.~\eqref{eq:example1} has the same concurrence as the Werner state. Therefore, we consider the Werner state to be our next application. In Fig.~\ref{fig:Werner} 
numerical results are presented for the average cost function $\bar{f}$ and the overall success probability which exhibit the same behavior found for the one-parameter family of states in Eq.~\eqref{eq:example1}. 
In this case, one can note an improvement of $0.04$ in $\bar{f}$ after $N=3$ iterations. This is less than the previously obtained value of $0.09$ shown in Fig.~\ref{fig:TWerner}.
Furthermore, this is accompanied by another numerical 
inaccuracy: The overall success probability at $\mathcal{C}=1$ is less than one. Given these results, the proposed method can find optimal entangling two-qubit gates for at least three iterations. It is also clear 
from these tests that the algorithm is always reducing $\bar{f}$, but from $N=3$ iterations this might not be an optimal improvement of the concurrence. This originates from the fact that the gradient $\nabla f (\boldsymbol{\alpha})$ [see Eq.~\eqref{eq:f}] is almost flat in the neighborhood of $\mathcal{C}=1$ for $N>2$ iterations and thus the numerical search for the optimal gate, i.e, 
the search for $\boldsymbol{\alpha}^* \in \mathbb{R}^{15}$, becomes inefficient.  

\begin{figure*}[t!]
 \begin{center}
 \includegraphics[width=.39\textwidth]{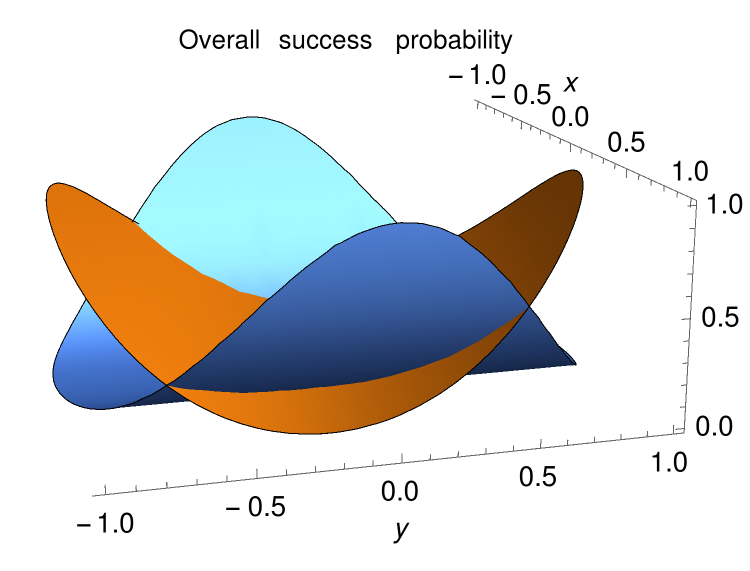}
\caption{Overall success probability of the state in Eq.~\eqref{eq:qrstate} after $N=3$ iterations as a function of $x$ and $y$. Both surfaces are similar in appearance. The one having maxima at $x=1$ 
belongs to the CNOT-based protocol and the other one having maxima at $y=1$ is obtained via the optimized numerical protocol.}
   \label{fig:XYos}
 \end{center}
\end{figure*}

Now, to test our method further, let us consider another state whose entanglement purification procedure is known. For this purpose we  
note that the state of 
{\it Example II.2.} in Eq.~\eqref{eq:example2} can be transformed using a separable gate
$b\otimes b^\dagger$,with $b$ defined in Eq.~\eqref{eq:localtr}, 
into the state
\begin{equation}
x\ketbra{\Psi^-}{\Psi^-}+(1-x)\ketbra{\Upsilon}{\Upsilon}, \label{eq:MaZs}
\end{equation}
with $x \in [0,1]$ and 
\begin{equation}
 \ket{\Upsilon}=\frac{1}{\sqrt2} \left(\ket{\Psi^+}+i\ket{\Phi^-}\right). \nonumber
\end{equation}
Using the unitary transformation  $b^\dagger \otimes b$ on Eq.~\eqref{eq:MaZs}, one obtains the state in Eq.~\eqref{eq:example2} which can be purified in one iteration using bilateral CNOT gates. For this reason,
it is expected that the optimal two-qubit gate that purifies the states in Eq.~\eqref{eq:MaZs} is the one that
achieves the task
in one iteration and given by a CNOT combined with single qubit gates in the form of $b$.
As both states are connected via a separable gate, they have the same concurrence $\mathcal{C}(x)=x$. However, it is important to note that the CNOT gate is not optimal for the state in Eq.~\eqref{eq:MaZs}, because after one iteration round
\begin{equation}
 \mathcal{C}'_{\text{CNOT}}(x)=\begin{cases} 0, & x \in [0,0.5] \\ 2x, \frac{2x-1}{1+x^2} & x \in (0.5,1], \end{cases} \nonumber
\end{equation} 
and $\mathcal{C}'_{\text{CNOT}}(x) \leqslant x$. The success probability is
\begin{equation}
 P_{\text{CNOT}}=\frac{1+x^2}{2}. \nonumber
\end{equation}
Thus, the CNOT gate without the non-symmetrical local transformation of Eq.~\eqref{eq:localtr} impairs the concurrence. In contrast to this analytical observation, numerical analysis with a uniform PDF yields already in the first iteration for both states an average cost function $\bar{f}\approx 0.0002$. To demonstrate the robustness of the numerical approach we provide examples of three non-uniform PDFs for $N=1$ 
iteration. First, we consider
\begin{equation}
p(x)=2x,\quad \text{with} \quad x \in [0,1], \nonumber
\end{equation}
which describes a situation, where states with higher concurrences are more likely to be subject to the purification. The resulting average cost function is $\bar{f}\approx 0.000004$. Second, we take 
\begin{equation}
p(x)=2(1-x) \quad \text{with} \quad x \in [0,1], \nonumber
\end{equation}
which puts more weight on states with low concurrences and find $\bar{f}\approx 0.0005$. Finally, we investigate a PDF
\begin{equation}
p(x)=6x(1-x) \quad \text{with} \quad x \in [0,1], \nonumber
\end{equation}
i.e., the states around the concurrence $\mathcal{C}(x)=0.5$ are more likely to participate in the purification, and obtain $\bar{f}\approx 0.00005$. These results demonstrate the effectiveness of our approach and up to a numerical precision these one-parameter families of states can be purified in one iteration. \newline
\begin{figure*}[t!]
 \begin{center}
 \includegraphics[width=.39\textwidth]{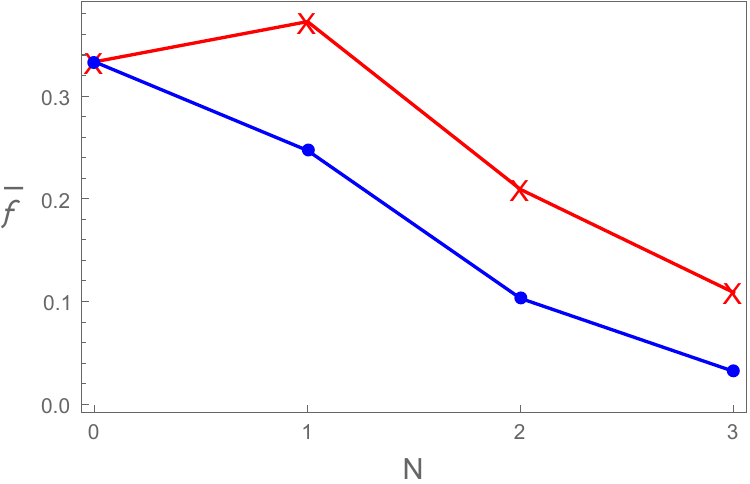}
 \includegraphics[width=.39\textwidth]{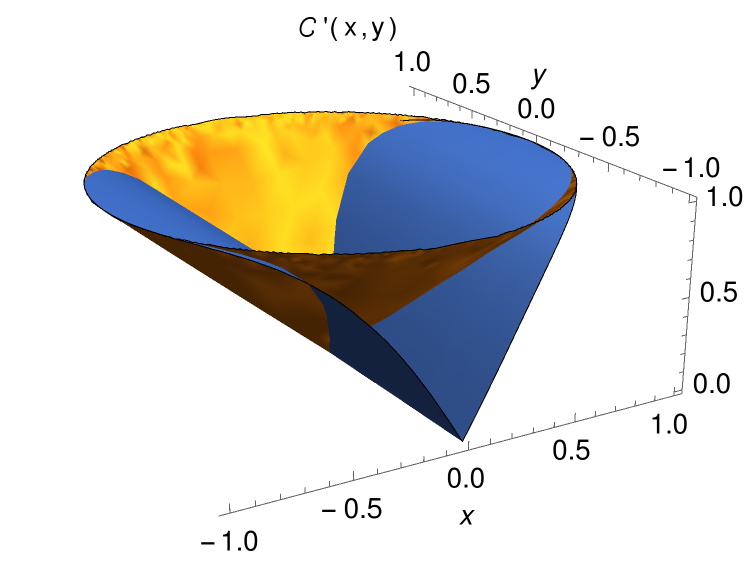}
\caption{Optimized purification protocol for the state in Eq.~\eqref{eq:qrstate} for 1000 parameter samples. Left panel: Average cost function $\bar{f}$ with a uniform PDF as a function of $N$, the number of iterations. 
Crosses are the results of the CNOT gate, whereas circles display the numerical optimization. They have been connected by lines to guide the eye. Right panel: Concurrence $\mathcal{C}'$
as a function of $x$ and $y$ after one purification round. The cone-type surface is obtained with our approach. A less optimal surface with minima at $x=0$ and along the $y$ axis is the result of the purification protocol 
with the CNOT gate. The sample size has been set to $M=1000$.}
   \label{fig:XY}
 \end{center}
\end{figure*}
Next we consider the following two-parameter family of states arising from a generation of distant entanglement in the context of a hybrid quantum repeater 
\cite{Bernad2},
\begin{equation}
\rho(x,y) = \frac{1}{4}
\left(
\begin{array}{cccc}
 1-x & i y & -i y & x-1 \\
 -i y & x+1 & -x-1 & i y \\
 i y & -x-1 & x+1 & -i y \\
 x-1 & -i y & i y & 1-x \\
\end{array}
\right). \label{eq:qrstate}
\end{equation}
Here $x,y \in \mathbb{R}$ and $x^2+y^2\leqslant1$. The concurrence of this state is $\sqrt{x^2+y^2}$. In order to relate the performance of our approach, we apply the CNOT gate based purification 
protocol to this family of states. We obtain
\begin{eqnarray}
 \mathcal{C}'_{\text{CNOT}}(x,y)&=&\frac{2|x|}{1+x^2}, \label{eq:xyCNOTc} \\
 P_{\text{CNOT}}&=&\frac{1+x^2}{2} \label{eq:xyCNOTp}
\end{eqnarray}
after $N=1$ and
\begin{eqnarray}
 \mathcal{C}'_{\text{CNOT}}(x,y)&=&\frac{4 |x| (1+x^2)}{1+6x^2+x^4}, \label{eq:xyCNOTc2} \\
 P_{\text{CNOT}}&=&\frac{1+6x^2+x^4}{2 (1+x^2)^2} \label{eq:xyCNOTp2}
\end{eqnarray}
after $N=2$ purification rounds. If one applies the unitary transformation in Eq.~\eqref{eq:localtr} on the state before the bilateral CNOT gates are performed, then the above results remain 
unchanged. These results demonstrate that the CNOT gate and the additional tricks, which have yielded optimal purifications for the one-parameter family of states, are not optimal in this scenario, since they are outperformed on average by our optimized protocol. 
Thus, the original CNOT-based purification protocols cannot exploit all the useful entanglement, because the concurrence 
and the success probability become independent of the $y$ variable. Using these analytical results and the uniform PDF
\begin{equation}
 p(x,y)=\frac{1}{\pi} \quad \text{with} \quad x^2+y^2 \leqslant 1, \nonumber
\end{equation}
we compare our approach with the above-presented analytical results [see Eq.~\eqref{eq:xyCNOTc} and Eq.~\eqref{eq:xyCNOTc2}]. In Fig.~\ref{fig:XY} we show that our optimized protocol provides after one iteration an $x$- and $y$-dependent concurrence. Although the 
numerically obtained concurrence is larger at $x=0$ and $y \in [-1,1]$ than the surface given by Eq.~\eqref{eq:xyCNOTc}, one can also observe the opposite at $y=0$ and $x \in [-1,1]$. However, the concurrence has more improvement with
our algorithm and the average cost function with the uniform PDF stays, for more iterations, lower than the original CNOT-based protocols [see the left panel in Fig. \ref{fig:XY}].      
In regard to the overall success probability, we let our algorithm work until the third iteration and in Fig.~\ref{fig:XYos} the results are compared with the CNOT-based purification protocols. The obtained 
surfaces differ only by a $\pi/2$ rotation around the $z$ axis. Therefore, there is not much difference in the overall success probability and from this aspect the CNOT gate can also be considered optimal, though 
it still cannot improve the concurrence as effectively as the gate obtained with our approach. However, one might think that with a proper choice of local unitary transformations, like the transformation 
in Eq.~\eqref{eq:localtr} used for Werner and one-step purifiable states, the application of the CNOT gate may result in an optimal purification protocol. Let us consider then two general local unitary 
transformations at locations $A$ and $B$ for the preparation of two-qubit states. It is enough to consider unitary transformations in $SU(2)$, for which the Euler angle parametrization reads \cite{Tilma2}
\begin{eqnarray}\label{eq:singleU}
 U_A&=& e^{i \sigma_z \alpha_1} e^{i \sigma_y \alpha_2} e^{i \sigma_z \alpha_3}, \\
 U_B&=& e^{i \sigma_z \beta_1} e^{i \sigma_y \beta_2} e^{i \sigma_z \beta_3},
\end{eqnarray}
where $\sigma_y$ and $\sigma_z$ are the Pauli matrices. Here $\alpha_1, \beta_1 \in [0, \pi]$, $\alpha_2,\beta_2 \in [0, \pi/2]$, and $\alpha_3, \beta_3 \in [0, 2\pi]$ are the Euler angles for  $SU(2)$. For the first
iteration, we analyze the output
unitary of the optimized protocol with \textit{Mathematica} \cite{CAS} and observe that the angles $\alpha_1$ and $\beta_1$ do not affect the effectiveness of the CNOT-based purification protocol. Optimal concurrences for the remaining four angles yield either the result in Eq.~\eqref{eq:xyCNOTc} or
\begin{equation}
 \mathcal{C}'_{\text{CNOT}}(x,y)=\frac{2|y|}{1+y^2},  
\end{equation}
with success probability
\begin{equation} 
 P_{\text{CNOT}}=\frac{1+y^2}{2}, 
\end{equation}
for $\alpha_2=\frac{\pi}{8}$, $\beta_2=\frac{3 \pi}{8}$, and $\alpha_3=\beta_3=\frac{3\pi}{4}$. For these results the average cost function with a uniform PDF yields $\bar{f}\approx 0.372$, which is still lower than
the average concurrence found by our approach. These results demonstrate that even with local unitary transformations the CNOT gate is globally not optimal for all values of $x$ and $y$. 
However, when $x$ and $y$
are known beforehand, 
one can combine the CNOT gate together with
the local unitary transformation
\begin{eqnarray}
 U^\dagger_A \otimes U_B, \quad \text{with} \quad U=\cos(\theta)\mathbb{I}_2 + \sin(\theta) \sigma_x, \label{eq:statedeptr}
\end{eqnarray}
where the angle $\theta$ is a function of $x$ and $y$. For example, for $x,y \geqslant 0$, we have
\begin{equation}
 \cos(\theta)=\sqrt{\frac{1}{2}+\sqrt{\frac{x+\sqrt{x^2+y^2}}{8 \sqrt{x^2+y^2}}}}. \nonumber
\end{equation}
Using this transformation before the purification protocol, we obtain the state
\begin{equation}
\frac{1+ \sqrt{x^2+y^2}}{2} \ketbra{\Psi^-}{\Psi^-}+\frac{1- \sqrt{x^2+y^2}}{2} \ketbra{\Phi^-}{\Phi^-}. \label{eq:Spec}
\end{equation}
Now, together with the single qubit gates in Eq.~
\eqref{eq:statedeptr},
the CNOT-based purification yields, after one iteration, the concurrence
\begin{equation}
 \mathcal{C}'_{\text{CNOT}}(x,y)=\frac{2\sqrt{x^2+y^2}}{1+x^2+y^2}, 
\end{equation}
with success probability
\begin{equation}
 P_{\text{CNOT}}=\frac{1+x^2+y^2}{2}. 
\end{equation}
It is immediate that $\mathcal{C}'_{\text{CNOT}}(x,y) \geqslant \sqrt{x^2+y^2}$, i.e., we are improving the concurrence. In this particular case, we do not need the average cost function in the numerical search, 
because the state is fixed. Our algorithm running with a single state with parameters $(x,y)$ as defined in Eq.~\eqref{eq:qrstate} instead of an ensemble of states can optimally improve the concurrence, but is unable to find this particular optimal solution presented above, because the transformation
\eqref{eq:statedeptr} together with CNOT gates is a nonsymmetrical transformation at nodes $A$ and $B$.

\begin{figure*}[ht!]
 \begin{center}
 \includegraphics[width=.45\textwidth]{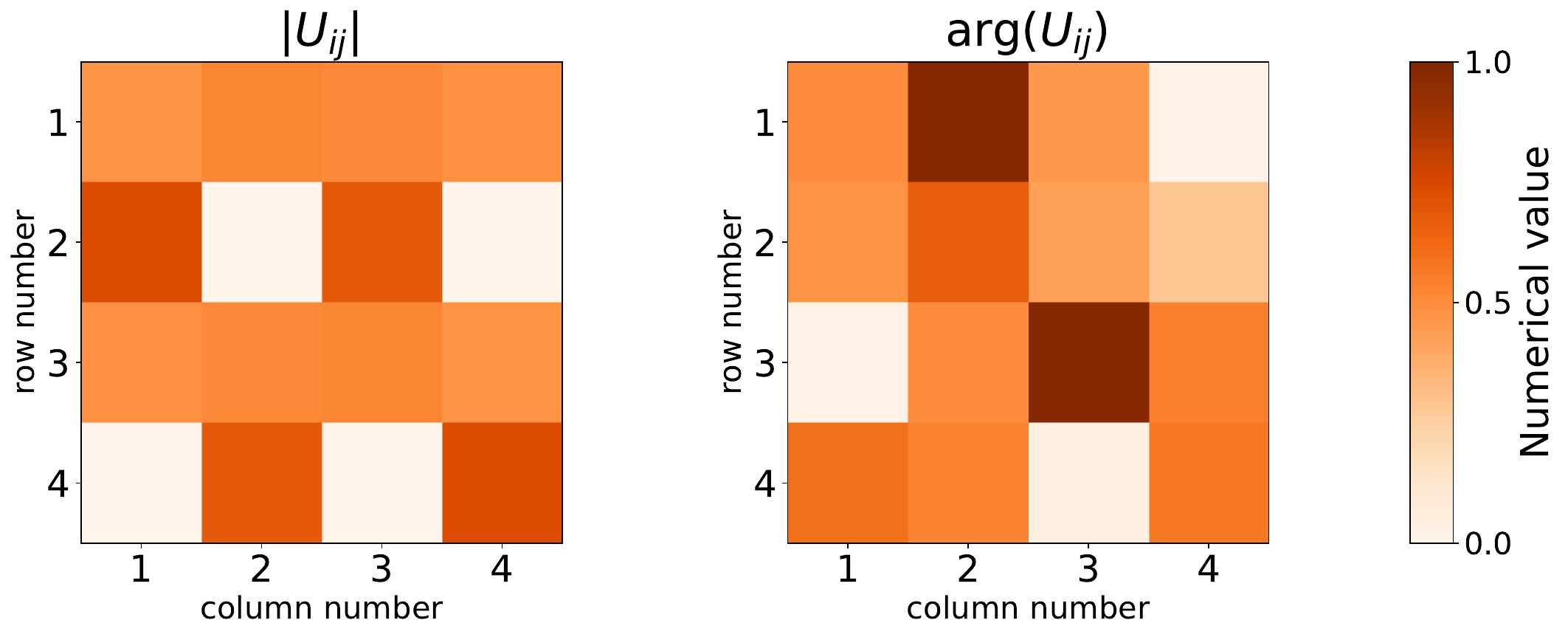}
 \hspace{1 cm}
 \includegraphics[width=.45\textwidth]{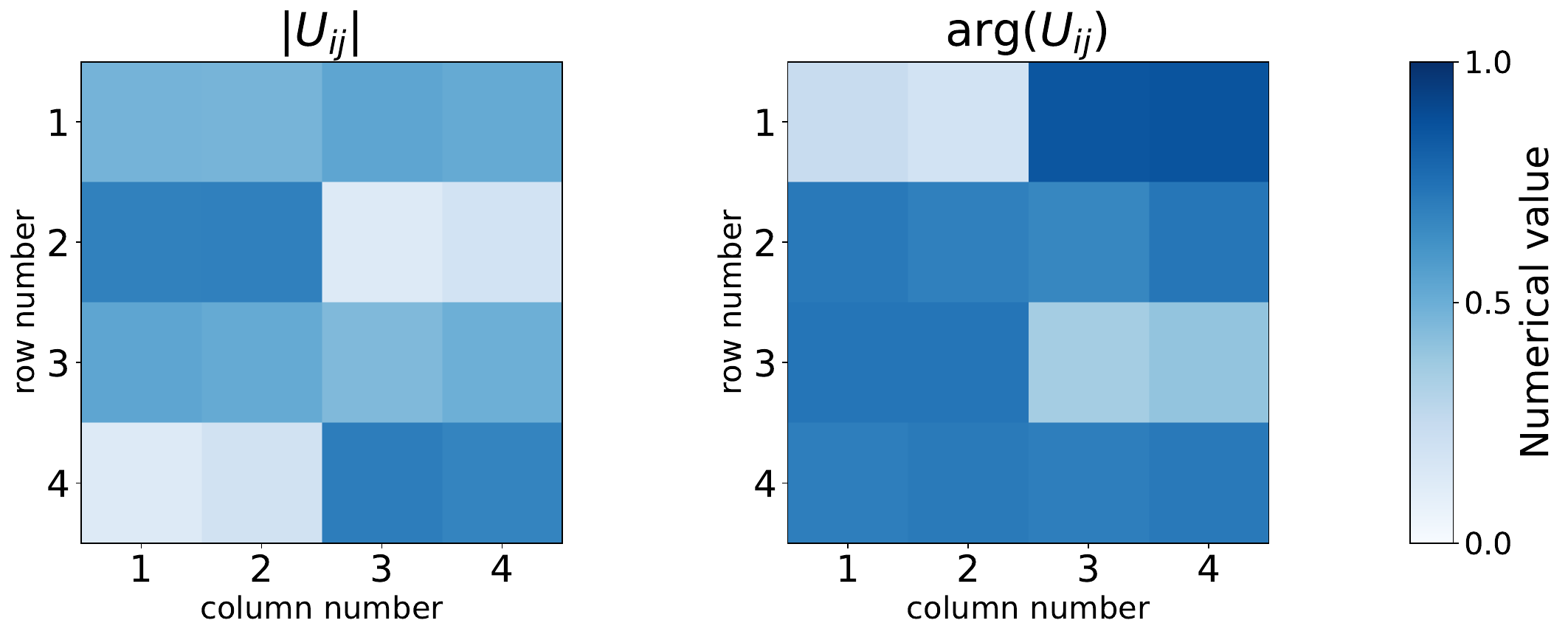}
\caption{Optimized unitaries  $U_N(\boldsymbol{\alpha}^*)$ -- the Euler angle parametrization in Eq.~\eqref{eq: eulerU} (left two color plots) -- and $\tilde{U}_N(\boldsymbol{\alpha}'^*)$ -- the CNOT circuit given in Eq.~ \eqref{eq:cnot_circuit} (right two color plots) -- for $N=1$. The optimal protocols are obtained for the state given in Eq.~\eqref{eq:qrstate} with $M=1000$ and the procedure discussed in Algorithm \ref{alg:ouralgo}. For each unitary, the first plot represents the absolute value of the matrix entries taken in the computational basis, whereas the second one represents the phase of the same entries divided by $2\pi$.
The color map describes values varying from 0 (white) to 1 (dark orange and dark blue).}
   \label{fig:XYgates}
 \end{center}
\end{figure*}

To show that optimization becomes more 
effective with an increased number of angles, we consider a less general unitary gate than the one in Eq.~\eqref{eq: eulerU}. This gate consists of one CNOT gate and four local unitary transformations of Eq.~\eqref{eq:singleU}
\begin{eqnarray} \label{eq:cnot_circuit}
    &\tilde{U}(\boldmath{\alpha}') = \big[ U_1(\alpha'_1, \alpha'_2, \alpha'_3) \otimes U_2(\alpha'_4, \alpha'_5, \alpha'_6) \big] \\ & \nonumber U_{\text{CNOT}} \big[ U_3(\alpha'_7, \alpha'_8, \alpha'_9) \otimes U_4(\alpha'_{10}, \alpha'_{11}, \alpha'_{12}) \big], \label{eq: 12U}
\end{eqnarray}
i.e., the quantum gate has a clear quantum circuit representation. Now $\boldmath{\alpha}'$ is a $12$-dimensional vector of the angles. We have inserted this gate into our algorithm and observe that
the results are almost as good as the optimization of a general gate with $15$ parameters (see Table \ref{tab:my_label}). Furthermore, the two optimal gates after $N=1$ iteration are presented in Fig. \ref{fig:XYgates}.

\begin{table}[]
    \centering
        \begin{tabular}{ |c|c|c| } 
         \hline
         N & $U(\boldsymbol{\alpha}^*)$ & $\tilde{U}(\boldsymbol{\alpha}'^*)$ \\ 
         \hline
         0 & 0.666 & 0.666 \\
         1 & 0.753 & 0.751 \\
         2 & 0.897 & 0.798 \\ 
         3 & 0.968 & 0.937 \\ 
         \hline
        \end{tabular}
    \caption{Comparison of the average concurrences produced with different parametrizations of unitary matrices given in Eqs.~\eqref{eq: eulerU} and \eqref{eq: 12U} for $M=1000$. The values in columns $2$ and $3$ represent the two best sequences found by the algorithm among ten different runs. We observe that the two different parametrizations provide us with improving concurrences. Nonetheless, the general two-qubit gate seems to perform slightly better.}
    \label{tab:my_label}
\end{table}

Finally, let us point out that our approach does not take into account operational or memory errors. These always depend on the implementation and if an experiment can provide us models for the errors then 
our approach can be easily extended. Another important experimental input is the PDF $p(\boldsymbol{x})$, which is usually subject to the method of generating entangled states between locations A and B. Furthermore, 
this PDF is assigned to the process of choosing a value $\boldsymbol{x}$ as a random event, i.e, we have the same two two-qubit pairs parametrized by $\boldsymbol{x}$ before the purification protocol.

In effect, we integrate away the parameter dependence of the states and thus our approach yields an optimal two-qubit gate on average. If the value of $\boldsymbol{x}$ is fixed in an experimental design, then our method 
provides an optimal two-qubit gate designed for this particular state. The optimal two-qubit gates can always be realized by three
CNOT gates and additional single-qubit gates \cite{Vidal,Vatan}, and therefore the experimental generation of this gate is possible with high fidelity \cite{Debnath}. Also in the context of trapped ions or superconducting quantum circuits, the generation of two-qubit entangling gates can be achieved with high precision using the M\o{}lmer-S\o{}rensen gate \cite{MS} or the $\sqrt{i \text{SWAP}}$ \cite{Fan} gate. Since two-qubit gates have a straightforward implementation in many physical settings, due to quantum compilation \cite{Martinez}, we argue that quantum computing and communication platforms could actually benefit from globally optimal gates for entanglement purification.

\section{Conclusions}
\label{Conclusions}

In the context of entanglement purification and recurrence protocols, we have presented a method to obtain optimal protocols. This method searches for the optimal two-qubit gate, which is applied bilaterally
at the nodes $A$ and $B$ in order to distill from an ensemble of mixed entangled pairs a higher fidelity state with respect to a maximally entangled state. Here we assumed that the same copies of the states can be generated before the purification protocol takes place, but a different experimental run could result in different states. Errors originating from local operations, memory requirements, or even 
classical communication have been neglected for now.

We numerically demonstrated the optimality of our proposal for several states. In the case of the Werner state, we found that several optimal two-qubit gates and their performances are the same as in the CNOT-based Deutsch protocol \cite{Deutsch}. Thus, for Werner states the optimality cannot be improved beyond the already known performance. Next we investigated a family of states, which can be purified 
in one step, i.e., two copies of mixed entangled states are enough to obtain a maximally entangled state. Here we immediately obtained a minimal average cost function $\bar{f} \approx 0$, as expected. Finally, we considered a two-parameter family of states which is obtained in theoretical models of a quantum repeater \cite{Bernad2}. Our numerical investigation demonstrated that a
single bilaterally applied CNOT gate cannot be globally optimal for these states. On the other hand, when the state is known beforehand, then parameter-dependent, nonsymmetric local transformations allow again the CNOT gate to be also one of the optimal two-qubit gates. This case exemplifies the difference between protocols that are optimal for a full class of parametrized states and those that are only optimal for a single state. We also investigated with our algorithm a concrete quantum circuit consisting of four different single-qubit gates and a CNOT gate, which is only a subset of the $SU(4)$ group. We found that the optimal two-qubit gate among all elements of this quantum circuit seems to be slightly worse than the one found among all matrices in $SU(4)$. This suggests that the search after an ensemble of optimal two-qubit gates can benefit from a general parametrization.

In conclusion, we have proposed a general method to optimize entanglement purification for an arbitrary family of states. Our algorithm \cite{Francesco} can find the two-qubit gates that on average induce the highest increase of entanglement. We remark that the method is general for the set of parameters defining the states. Optimizing for a single state is also possible, as shown in one of the 
presented examples. Furthermore, we focused on the degree of entanglement measured by the concurrence and not on the fidelity with respect to a particular Bell state. 
This is motivated by the fact that a general 
entanglement purification protocol may not always purify towards a given Bell state, but rather a 
maximally entangled state. As we optimize among many
different protocols, it is necessary to use a 
measure for which all maximally entangled states are equivalent.
A possible drawback is that one cannot know with certainty the final maximally entangled state produced by the protocol. This and other issues such as different input states, nonsymmetric two-qubit gates, and nonideal local operations remain open questions. However, this work aims to introduce a concept of globally optimized recurrence protocols that is flexible enough to incorporate the 
aforementioned points in further investigations.

\begin{acknowledgments}
We thank Michael Schilling for his help with the plots. This work was supported by the DFG under Germany's Excellence Strategy, Cluster of Excellence Matter and Light for Quantum Computing (ML4Q) Grant No. EXC 2004/1-390534769, the  German  Ministry for  Education  and  Research,  under  QSolid,  Grant  no.~13N16149, and AIDAS - AI, Data Analytics and Scalable Simulation - which is a Joint Virtual Laboratory gathering the Forschungszentrum J\"ulich (FZJ) and the French Alternative Energies and Atomic Energy Commission (CEA). The results were obtained using the libraries JAX \cite{JAX}, Scipy \cite{Scipy} and QuTiP \cite{Qutip}.

\end{acknowledgments}

\appendix
\begin{widetext}
\section{Gell-Mann type basis}
\label{AppendixA}

In this appendix, details concerning the Gell-Mann type basis for the Lie algebra of $SU(4)$ are shown. The matrices read
\begin{eqnarray}
 \sigma_1 &=& \begin{pmatrix} 0 & 1 & 0 & 0 \\ 1 & 0 & 0 & 0 \\ 0 & 0 & 0 & 0\\ 0 & 0 & 0 & 0  \end{pmatrix}, \quad \quad \quad \,\,\,
 \sigma_2 = \begin{pmatrix} 0 & -i & 0 & 0 \\ i & 0 & 0 & 0 \\ 0 & 0 & 0 & 0 \\  0 & 0 & 0 & 0  \end{pmatrix}, \quad \quad \quad \,\,\,
 \sigma_3 = \begin{pmatrix} 1 & 0 & 0 & 0 \\ 0 & -1 & 0 & 0 \\ 0 & 0 & 0 & 0 \\ 0 & 0 & 0 & 0 \end{pmatrix}, \nonumber \\
\sigma_4 &=& \begin{pmatrix} 0 & 0 & 1 & 0 \\ 0 & 0 & 0 & 0 \\ 1 & 0 & 0 & 0 \\ 0 & 0 & 0 & 0 \end{pmatrix}, \quad \quad \quad \,\,\,
\sigma_5 = \begin{pmatrix} 0 & 0 & -i & 0 \\ 0 & 0 & 0 & 0 \\ i & 0 & 0 & 0 \\ 0 & 0 & 0 & 0 \end{pmatrix}, \quad \quad \quad \,\,\,
\sigma_6=\begin{pmatrix} 0 & 0 & 0 & 0 \\ 0 & 0 & 1 & 0 \\ 0 & 1 & 0 & 0 \\ 0 & 0 & 0 & 0 \end{pmatrix}, \nonumber \\
\sigma_7 &=& \begin{pmatrix} 0 & 0 & 0 & 0 \\ 0 & 0 & -i & 0 \\ 0 & i & 0 & 0 \\ 0 & 0 & 0 & 0  \end{pmatrix}, \quad 
\sigma_8= \frac{1}{\sqrt{3}} \begin{pmatrix} 1 & 0 & 0 & 0 \\ 0 & 1 & 0 & 0 \\ 0 & 0 & -2 & 0 \\ 0 & 0 & 0 & 0  \end{pmatrix}, \quad \quad \quad \,\,\, 
\sigma_9=\begin{pmatrix} 0 & 0 & 0 & 1 \\ 0 & 0 & 0 & 0 \\ 0 & 0 & 0 & 0 \\ 1 & 0 & 0 & 0 \end{pmatrix}, \nonumber \\
\sigma_{10} &=& \begin{pmatrix} 0 & 0 & 0 & -i \\ 0 & 0 & 0 & 0 \\ 0 & 0 & 0 & 0 \\ i & 0 & 0 & 0 \end{pmatrix}, \quad \quad \,\,\,\, 
\sigma_{11} = \begin{pmatrix} 0 & 0 & 0 & 0 \\ 0 & 0 & 0 & 1 \\ 0 & 0 & 0 & 0 \\ 0 & 1 & 0 & 0 \end{pmatrix}, \quad \quad \quad \,\,\,
\sigma_{12}=\begin{pmatrix} 0 & 0 & 0 & 0 \\ 0 & 0 & 0 & -i \\ 0 & 0 & 0 & 0 \\ 0 & i & 0 & 0 \end{pmatrix}, \nonumber \\
\sigma_{13} &=& \begin{pmatrix} 0 & 0 & 0 & 0 \\ 0 & 0 & 0 & 0 \\ 0 & 0 & 0 & 1 \\ 0 & 0 & 1 & 0 \end{pmatrix}, \quad \quad \quad \,\,\,
\sigma_{14} = \begin{pmatrix} 0 & 0 & 0 & 0 \\ 0 & 0 & 0 & 0 \\ 0 & 0 & 0 & -i \\ 0 & 0 & i & 0 \end{pmatrix}, \quad 
\sigma_{15}=\frac{1}{\sqrt{6}}\begin{pmatrix} 1 & 0 & 0 & 0 \\ 0 & 1 & 0 & 0 \\ 0 & 0 & 1 & 0 \\ 0 & 0 & 0 & -3 \end{pmatrix}.
\label{Gell-Mann}
\end{eqnarray}
Actually, these matrices together with the identity matrix 
\begin{equation}
  \sigma_0 =\begin{pmatrix} 1 & 0 & 0 & 0 \\ 0 & 1 & 0 & 0 \\ 0 & 0 & 1 & 0\\ 0 & 0 & 0 & 1  \end{pmatrix}, \nonumber
\end{equation}
form an orthogonal basis for the space $M_n(\mathbb{C})$ of $4 \times 4$ matrices with complex entries equipped with the Hilbert-Schmidt scalar product
\begin{equation}
 \langle A,B\rangle=\mathrm{Tr} \{A^\dagger B\},\quad A,B \in M_n(\mathbb{C}), \nonumber
\end{equation}
where $A^\dagger$ is the adjoint of $A$. We first note that $\sigma^\dagger_i=\sigma_i$ for all $i \in \{0,1,2, \dots, 15\}$. Therefore, every matrix $X \in M_n(\mathbb{C})$ can be written as
\begin{equation}
 X= \sum^{15}_{i=0} \frac{\langle \sigma_i ,X\rangle}{\langle \sigma_i, \sigma_i\rangle} \sigma_i,  \quad 
 X^\dagger= \sum^{15}_{i=0} \frac{\langle \sigma_i ,X\rangle^*}{\langle \sigma_i, \sigma_i\rangle} \sigma_i, \nonumber
\end{equation}
where $z^*$ is the complex conjugate of the complex number $z \in \mathbb{C}$. In this fashion, we can obtain another representation for any element $U \in SU(4)$ fulfilling $U^\dagger U =U U^\dagger =\sigma_0$
and including also that its determinant is equal to one. This is also a minimal parametrization, however a cumbersome one compared to the Euler angle parametrization. In general, a Gell-Mann-type basis for the Lie algebra of $SU(n)$ can always be obtained, see Refs. \cite{Tilma2,Bertini}.    
\end{widetext}

\end{document}